\documentclass[
  aps,
  prl,
  reprint,                
  superscriptaddress,
  amsmath,
  amssymb,
  longbibliography,nofootinbib
]{revtex4-1}              

\usepackage{graphicx}
\usepackage[caption=false]{subfig}
\usepackage{bm}
\usepackage{booktabs}
\usepackage[hidelinks]{hyperref}
\usepackage{tikz}
\usetikzlibrary{patterns}
\usepackage{multirow}
\usepackage{comment}

\def\Tr#1{\operatorname{Tr}\!\left(#1\right)}
\def\tr#1{\operatorname{tr}\!\left(#1\right)}

\def\Q{\mathcal{Q}}
\def\d{\mathrm{d}}

\def\rz{\mathsf{z}}
\def\ry{\mathsf{y}}
\def\rx{\mathsf{x}}

\begin{document}

\title{BPS Bootstrap}

\author{Yiming Chen}
\email{ymchen.phys@gmail.com} 
\affiliation{Department of Physics, Cornell University, Ithaca, NY 14853, USA}
\affiliation{Leinweber Institute for Theoretical Physics, Stanford University, Stanford, CA, 94305, USA}


\begin{abstract}
We apply the large $N$ matrix bootstrap technique to constrain expectation values of simple operators in heavy BPS states that are dual to classical geometries. For states dual to horizonless geometries, their explicit moduli give rise to large allowable regions in the bootstrap, while for black holes, a converging bootstrap should find islands. We explore this method in $\mathcal{N}=4$ $SU(N)$ super Yang-Mills theory at large $N$, working at weak coupling and focusing on BPS states with scaling dimensions of order $N^2$. For the $1/2$, $1/4$, and $1/8$-BPS sectors, the bootstrap finds large regions, in agreement with an eigenvalue description of such states in the limit of large charges. We also present preliminary results for a BMN truncation of the $1/16$-BPS sector, which contains states dual to both horizonless geometries and black holes. Isolating the black hole states in the bootstrap remains a challenge.
\end{abstract}

\maketitle

\section{Introduction}

An important question in gauge/gravity duality is to understand features of the black hole microstates in the gauge theory.
A flurry of recent activities has  made progress on this front. In the 4d $\mathcal{N}=4$ SYM theory, a class of ``fortuitous" BPS states whose BPS condition relies on finite $N$ effects was identified \cite{Chang:2022mjp,Choi:2022caq,Choi:2023znd,Budzik:2023vtr,Choi:2023vdm,Chang:2024zqi}. They differ from the ``monotone" multi-graviton states, whose BPS condition is independent of $N$, and account for the black hole entropy \cite{Cabo-Bizet:2018ehj,Choi:2018hmj,Benini:2018ywd}. The idea of fortuity has since been developed along several directions, including operators that describe ``dressed" black holes \cite{Choi:2023vdm,deMelloKoch:2024pcs}, generalization to different theories \cite{Chang:2024lxt,Chang:2025rqy,Hughes:2025tdy,Chen:2025sum,Chang:2025wgo,Belin:2025hsg,Choi:2025pqr,Behan:2025hbx,Tierz:2026vrs,Giusto:2026rpl,Belin:2026vfd,Komatsu:2026ano,Choi:2026faq,Shi:2026rbr,Berkooz:2026dbd}, connections to quantum chaos \cite{Chen:2024oqv,Chen:2026vml,Miyahara:2026iso} and possible extensions \cite{deMelloKoch:2025ngs,deMelloKoch:2025cec,Chryssanthacopoulos:2026mni,Choi:2026faq}. 

Much recent progress has been made by studying the cohomologies of the supercharge $\mathcal{Q}$, at leading order in the weak coupling expansion and at small values of $N$. To truly access black hole microstates in the gravity regime, one would however need to go to strong coupling and large $N$. Effects of loop corrections to $\mathcal{Q}$ have been recently studied in \cite{Gadde:2025yoa,Chang:2025mqp,Choi:2025bhi,Budzik:2025zvu,Kim:2026rnx,Kibe:2026qid}. In this paper, instead, we would try to make progress towards the large $N$ limit. Importantly, when we take $N\rightarrow \infty$, we also take the energies of the states to scale as $\mathcal{O}(N^2)$ so that they limit to classical geometries (with or without horizons) in the bulk. Such operators are also referred to as ``huge operators" and have been studied from other angles, see e.g. \cite{Kazakov:2024ald,Guerrieri:2025ytx,Anempodistov:2025maj}.

The bootstrap technique for matrix quantum mechanical systems \cite{Han:2020bkb,Lin:2023owt,Lin:2024vvg,Lin:2025srf,Laliberte:2025xvk} (``matrix bootstrap") is an effective technique that allows one to access information directly in the large $N$ limit. The prescription that we will adopt follows closely that of \cite{Lin:2023owt,Lin:2024vvg,Lin:2025srf}. Here we only highlight the main ingredients, leaving further details to the references and Appendix \ref{app:bootstrap}. In the bootstrap program, one constrains expectation values of single trace operators by demanding positivity of an inner product matrix
\begin{equation}\label{Mpositivity}
	\mathcal{M}_{ij} \equiv \langle \tr{\mathcal{O}_i \bar{\mathcal{O}}_j } \rangle\,, \quad \mathcal{M}\succeq 0\,
\end{equation}
where $\mathcal{O}_i, \mathcal{O}_j$ are strings formed by adjoint letters and their conjugates. The positivity constraint (\ref{Mpositivity}) is purely kinematical. One further supplements it with dynamical constraints that are specific to $\langle \cdot\rangle$, i.e. the states in which the expectation values are evaluated. 

In the literature, $\langle \cdot\rangle$ is often taken to denote ground state expectation values.\footnote{See however examples of using matrix bootstrap to access finite temperature and real time physics \cite{Fawzi:2023fpg,Cho:2024kxn,Cho:2025vws,Klebanov:2026rfc,Dodelson:2026hjm}.}
For a supersymmetric matrix quantum mechanical system where $H = \{\Q,\Q^\dagger\}$, the supersymmetric ground state is annihilated by both $\Q$ and $\Q^\dagger$ and this results in loop equations
\begin{equation}\label{Qloop}
	\langle \{\Q, \mathcal{O}\}  \rangle = 0
\end{equation}
for any single trace operator $\mathcal{O}$. The loop equations involving $\Q^\dagger$ are related by Hermitian conjugation. These equations reduce the number of free variables in the bootstrap program and lead to stronger bounds.
The starting observation for us is that exactly the same formulation applies to BPS states, since they are also annihilated by $\Q$ and $\Q^\dagger$. In the field theory examples that we will focus on, we have $\{\Q,\Q^\dagger\} \propto H - E_{\textrm{BPS}}$, where $E_{\textrm{BPS}}$ will be chosen to scale as $\mathcal{O}(N^2)$. This enables us to probe physics of classical geometries including black holes with essentially the same formalism.

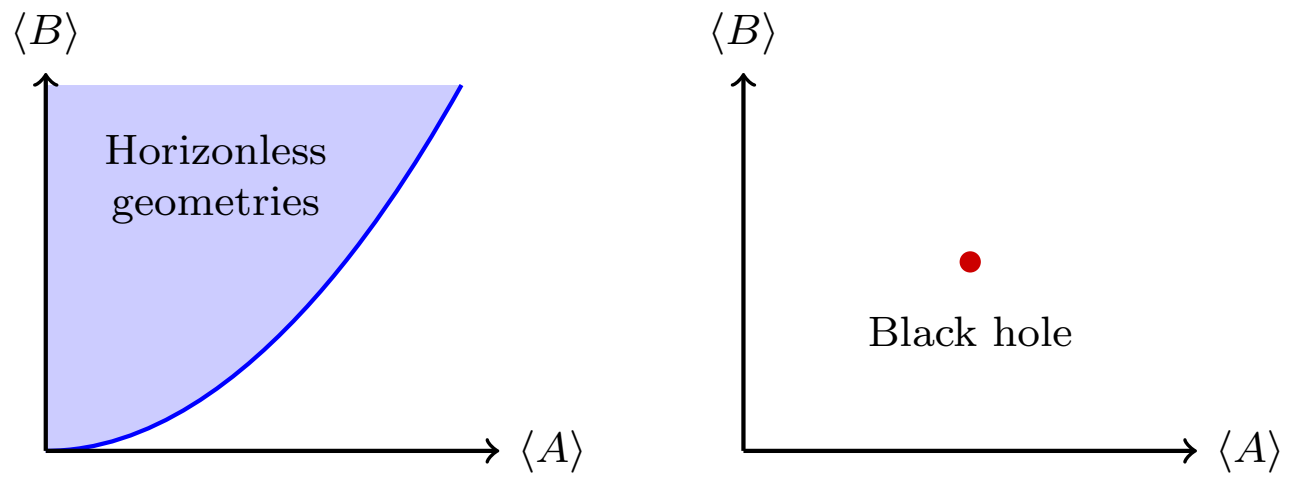
\begin{figure}[b!]
  \centering

  \begin{tikzpicture}[x=1.25cm, y=0.5cm, font=\footnotesize]
    \fill[blue!20]
      (0,0) parabola (2.2,4.84)
      -- (0,4.84) -- cycle;

    \draw[blue, thick]
      (0,0) parabola (2.2,4.84);

    \draw[->, thick] (0,0) -- (2.4,0)
      node[right] {$\langle A\rangle$};
    \draw[->, thick] (0,0) -- (0,5)
      node[above] {$\langle B\rangle$};

    \node[align=center] at (0.9,3.6)
      {Horizonless\\geometries};
  \end{tikzpicture}
  \hfill
  \begin{tikzpicture}[x=1.25cm, y=0.5cm, font=\footnotesize]
    \draw[->, thick] (0,0) -- (2.4,0)
      node[right] {$\langle A\rangle$};
    \draw[->, thick] (0,0) -- (0,5)
      node[above] {$\langle B\rangle$};

    \fill[red!80!black] (1.2,2.5) circle (2pt);
    \node[below=7pt] at (1.2,2.5) {Black hole};
  \end{tikzpicture}

  \caption{We sketch our expectation assuming infinite power of the bootstrap. (a) For BPS sectors that are described by horizonless geometries with explicit moduli in the large $N$ limit, bootstrap should find a continent of different allowed values. (b) On the other hand, bootstrap should find an island when the BPS states are described by a supersymmetric black hole solution.}
  \label{fig:horizon}
\end{figure}

Despite the similar formalism, there are important new features that arise in the BPS bootstrap. Instead of having a single or a few ground states, the number of BPS states is vast, often growing as $e^{\mathcal{O}(N^{\alpha})}$ with $\alpha>0$. In such situations, $\langle \mathcal{O}\rangle$ does not mean the expectation value of the operator $\mathcal{O}$ in a particular BPS state, but rather 
\begin{equation}
	\langle \mathcal{O} \rangle \equiv \frac{ \textrm{Tr}\left[\rho_{\textrm{BPS}} \mathcal{O}\right]}{\textrm{Tr}\left[\rho_{\textrm{BPS}} \right]}\,,
\end{equation}
where $\rho_{\textrm{BPS}}$ can represent an arbitrary mixed state in the BPS subspace, which remains unspecified in the bootstrap program. 
Of course,  one is free to impose further conditions on $\rho_{\textrm{BPS}}$, such as focusing on the states in a definite charge sector (up to $o(N^2)$ accuracy) by fixing the values of conserved charges in the bootstrap.\footnote{Consider a single charge $j$. Fixing the value of $\langle j \rangle = j_0$ in the bootstrap does not necessarily mean that the BPS states have charge $j_0$, as one can also achieve the same expectation value with a classical mixture of states with $j_1, j_2$ satisfying $j_1 < j_0 < j_2$. We will further impose large $N$ factorization $\langle j\mathcal{O}\rangle =j_0\langle\mathcal{O}\rangle$, including $\langle j^2\rangle = \langle j\rangle^2$ that requires $j$ to have zero variance and removes the latter possibility.} However, even with all charges fixed, there is usually still a large number of BPS states.

In fact, having a large number of BPS states leads to an interesting aspect of the problem. Often in bootstrap, one wishes for a convergence of the allowed values into tinier ``islands" as one increases the truncation level. Here, since  the large number of  BPS states can each have distinctive expectation values, a converging island should only come as a surprise. The typical situation should instead be that of a non-shrinking region, or as we will see, a ``continent" that can be unbounded in certain directions. As we will show in examples, the continent reflects that even within the same charge sector, the BPS states are dual to different horizonless geometries. 

When can we possibly hope to find a converging island? It can occur when the BPS states are described by a supersymmetric \emph{black hole}.\footnote{Since our analysis is at weak coupling of the Yang-Mills theory, the black holes are highly stringy. Nonetheless, the Schwarzian mode is expected to dominate close to extremality \cite{Maldacena:2016hyu}.} Lin, Maldacena, Rozenberg and Shan (LMRS) \cite{Lin:2022rzw,Lin:2022zxd} showed that the super-Schwarzian theory \cite{Stanford:2017thb} implies $\langle\mathcal{O}\rangle$ can at most differ by $\sim S^{-\Delta}\rightarrow 0$ for different choices of $\rho_{\rm BPS}$, where $S \sim N^2$ is the entropy of the black hole and $\Delta$ is the AdS$_2$ dimension of the operator $\mathcal{O}$.  From the boundary perspective, the statement that all BPS states have the same expectation value of simple operators in the large $N$ limit is a special version of the eigenstate thermalization hypothesis (ETH) \cite{Srednicki:1994mfb}, where a simple operator projects into a random matrix in the BPS subspace, whose  eigenvalues span over a width $\sim S^{-\Delta}$. From a bootstrap perspective, however, the distinction between microstate geometries that have continuous moduli and black holes appears quite non-trivial. We illustrate this distinction in Figure \ref{fig:horizon}.

We should immediately emphasize that despite such an expectation, we \emph{have not} yet found islands associated with black holes. In fact, we have not been able to isolate black holes at all.  The basic difficulty would appear to be that for the BMN sector (to be introduced shortly) that is expected to include black hole states, it also simultaneously contains non-black hole states.  The non-black hole states ``contaminate" the bounds for the black hole states, and we have not found a decisive way to separate them. We will nonetheless present some preliminary explorations, with the hope that further improvements on the numerics can guide us to isolate the black hole. 
 
In addition to our paper, we also provide an interactive toy bootstrap program \cite{BPSBootstrapCode}, in which one has the freedom of specifying their favorite supercharge $\Q$ and exploring the properties of corresponding BPS states.

\section{Setup}

Our goal is to apply the matrix bootstrap method to the $1/16$-BPS sector of the 4d $\mathcal{N} = 4$ SYM theory with $SU(N)$ gauge group. Putting the theory on $S^3\times \mathbb{R}_t$, states are labeled by the two Cartans $J_1, J_2$ of the $SO(4)$ rotation on $S^3$, as well as the three Cartans $R_1, R_2, R_3$ of the $SO(6)_R$ R-symmetry group. The $1/16$-BPS states are those whose energies saturate the BPS bound \cite{Kinney:2005ej}
\begin{equation}
	E \geq E_{\textrm{BPS}} =   R_1 + R_2 +R_3+ J_1 + J_2\,.
\end{equation}
One can build up $1/16$-BPS states in the free theory by starting from a set of BPS letters. By turning on interactions, most of them are lifted and the remaining ones will be our focus. To access the full $1/16$-BPS sector, one needs to include infinitely many letters. To make the problem amenable to matrix bootstrap, we follow \cite{Choi:2023vdm} and consider the ``BMN truncation" where only seven letters are kept
\begin{equation}\label{letters}
	\phi^1,\,\phi^2,\,\phi^3,\,\psi_1,\,\psi_2,\,\psi_3,\,f\,.
\end{equation}
We summarize the charges of these letters in Appendix \ref{app:BMN}.
With a slight abuse of notation, we will take $\phi^1,\phi^2,\phi^3,f$ and $\psi_1,\psi_2,\psi_3$ to denote bosonic and fermionic matrix \emph{creation} operators, while the barred variables denote \emph{annihilation} operators.\footnote{This is different from the normal field theory notation where the letters denote complex fields. We further work in a normalization in which these letters are related to the canonical normalization by a factor of $\sqrt{N}$, i.e. $\phi = \frac{1}{\sqrt{N}} \Phi$, where $[\bar{\Phi}^{i}_{\,j}, \Phi^k_{\,l}] = \delta^{i}_{\,l}\delta^{k}_{\,j} - \frac{1}{N} \delta^{i}_{\,j} \delta^{k}_{\,l}. $ Similarly, $\textrm{tr}$ is related to an ordinary trace $\textrm{Tr}$ by $\textrm{tr}(\cdot)=\frac{1}{N}\textrm{Tr}(\cdot) $. }  The BMN truncation is closed at one-loop, in the sense that the action of the supercharges $\Q,\Q^\dagger$ does not introduce new letters. Explicitly, 
up to an overall normalization constant that will be unimportant for our study, we have\footnote{The additional factor of $\sqrt{2}$ compared to \cite{Choi:2023vdm} comes from different normalizations of the gauge field. Here we choose to normalize $f$ in the same way as $\phi^i$, so they could be treated democratically in the bootstrap.}
\begin{equation}
\mathcal{Q}
= -\epsilon_{ijk}\tr{[\phi^j,\phi^k]\bar{\psi}_i}
+\sqrt{2}\tr{[\phi^i,\psi_i]\bar{f}}\,.
\end{equation}

Despite the truncation, the BMN sector already exhibits fortuitous states  and an $\mathcal{O}(N^2)$ entropy that hints towards black hole-like physics \cite{Choi:2023vdm,Chang:2024lkw}. Therefore, we believe it is a good playground to explore the matrix bootstrap technique. 
Even though our analysis is at weak couplings, there is a sense that the problem is strongly coupled due to the infinite Euclidean evolution needed to project into the BPS states. For this reason, the expectation values in one-loop BPS states are not easily accessible by ordinary perturbative methods, making the bootstrap method particularly valuable.

Before we move on to the results, let us also spell out some of the additional assumptions on the BPS states in our bootstrap program. 
\begin{itemize}
	\item We assume that $\rho_{\textrm{BPS}}$ in which the expectation values are evaluated has  definite charges, so that $\langle \mathcal{O}\rangle \neq 0$ only when $\mathcal{O}$ is neutral.
	\item We assume large $N$ factorization holds in $\rho_{\textrm{BPS}}$. This is reasonable if the state is described by a single classical geometry in the bulk. 
\end{itemize}

\section{The Half, Quarter, Eighth BPS Sectors}

Before we consider the full BMN sector, it is worthwhile to consider some simpler subsectors as warm-ups. Our goal is not to be thorough, but only to highlight some main features. 
There are several more supersymmetric sectors of $\mathcal{N}=4$ SYM theory, which are known to not contain black holes.  Luckily, we do not need to introduce them one by one, as we can arrive at them by imposing further restrictions on the BMN sector:
\begin{itemize}
	\item $\frac{1}{2}$-BPS: only include $z \equiv \phi^3$\,,
	\item $\frac{1}{4}$-BPS: include $x \equiv \phi^1$ and $z$\,,
	\item $\frac{1}{8}$-BPS: include $x,z$ and $y\equiv \phi^2$.
\end{itemize}
Let's start with the simplest case of $1/2$-BPS states. The supercharge $\mathcal{Q}$ is trivially zero in this sector, so all the non-trivial bounds come from positivity.  For instance, demanding the positivity of the $2\times 2$ block formed by the strings $1$ and $z\bar{z}$, 
\begin{equation}
	\mathcal{M} = \begin{pmatrix}  1 &  \langle\tr{z\bar{z}} \rangle \\
\langle	\tr{z\bar{z}}  \rangle & \langle \tr{z\bar{z}z \bar{z}} \rangle \end{pmatrix}\succeq 0\,,
\end{equation}
one finds that
\begin{equation}\label{z4z2}
	\langle \tr{z\bar{z}z \bar{z}} \rangle  \geq \langle\tr{z\bar{z}} \rangle^2\,
\end{equation}
One can verify that in this case, increasing the truncation level in the bootstrap does not yield a stronger bound, nor does it lead to an upper bound on $\langle \tr{z\bar{z}z \bar{z}} \rangle$.


\begin{figure}[b!]
\centering
\begingroup

\def\plotX{3.0}             
\def\plotY{1.1}             
\def\panelSize{1.00cm}
\def\dropletScale{0.13}    
\def\boxStroke{0.15pt}
\def\outerRingStroke{0.40pt}

\resizebox{0.9\columnwidth}{!}{%
\begin{tikzpicture}[
  x=1cm,y=1cm,font=\large,
  line cap=round,line join=round,>=latex,
  panel/.style={
    draw=black!60,line width=\boxStroke,fill=white,
    minimum size=\panelSize,inner sep=0pt,outer sep=0pt
  },
  connector/.style={
    black,line width=0.4pt,dash pattern=on 2pt off 1.7pt
  }
]

\begin{scope}[x=\plotX cm,y=\plotY cm]
  \fill[blue!14]
    (0,0) parabola (2.2,4.84) -- (0,4.84) -- cycle;

  \draw[blue!80!black,line width=0.75pt]
    (0,0) parabola (2.2,4.84);

  \draw[->,line width=0.6pt] (0,0) -- (2.4,0)
    node[right=3pt]
    {$\langle\operatorname{tr}(z\bar z)\rangle$};

  \draw[->,line width=0.6pt] (0,0) -- (0,5.05)
    node[above=3pt]
    {$\langle\operatorname{tr}(z\bar z z\bar z)\rangle$};

  \coordinate (vacuum-point) at (0,0);
  \coordinate (annulus-point) at (1.15,1.3225);
  \coordinate (two-annuli-point) at (1.15,4.350625);
\end{scope}

\node[panel] (vacuum-inset) at (-1.40,1.25) {};
\node[panel] (annulus-inset) at (4.80,0.85) {};
\node[panel] (two-annuli-inset) at (5.70,5.40) {};

\draw[connector]
  (vacuum-inset.south west) -- (vacuum-point)
  -- (vacuum-inset.north east);

\draw[connector]
  (annulus-inset.south west) -- (annulus-point)
  -- (annulus-inset.north east);

\draw[connector]
  (two-annuli-inset.north west) -- (two-annuli-point)
  -- (two-annuli-inset.south east);


\begin{scope}[
  shift={(vacuum-inset.center)},scale=\dropletScale
]
  \fill[black] (0,0) circle[radius=1];
\end{scope}

\begin{scope}[
  shift={(annulus-inset.center)},scale=\dropletScale
]
  \fill[black,even odd rule]
    (0,0) circle[radius={sqrt(2.15)}]
    (0,0) circle[radius={sqrt(1.15)}];
\end{scope}

\begin{scope}[
  shift={(two-annuli-inset.center)},scale=\dropletScale
]
  \fill[black,even odd rule]
    (0,0) circle[radius={sqrt(1.725)}]
    (0,0) circle[radius={sqrt(0.775)}];

  \path[
    fill=black,draw=black,
    line width=\outerRingStroke,even odd rule
  ]
    (0,0) circle[radius={sqrt(9.275)}]
    (0,0) circle[radius={sqrt(9.225)}];
\end{scope}

\foreach \point in {vacuum-point,annulus-point,two-annuli-point}
  \fill[red!80!black] (\point) circle[radius=1.5pt];

\end{tikzpicture}%
}
\endgroup

\caption{We sketch the simple bound in (\ref{z4z2}). The boundary of the allowed region ({\color{blue}{blue}}) corresponds to simple shapes of the droplet, such as a disk or an annulus, while the unbounded direction corresponds to moving parts of the droplet very far away.}
\label{fig:droplet}
\end{figure}
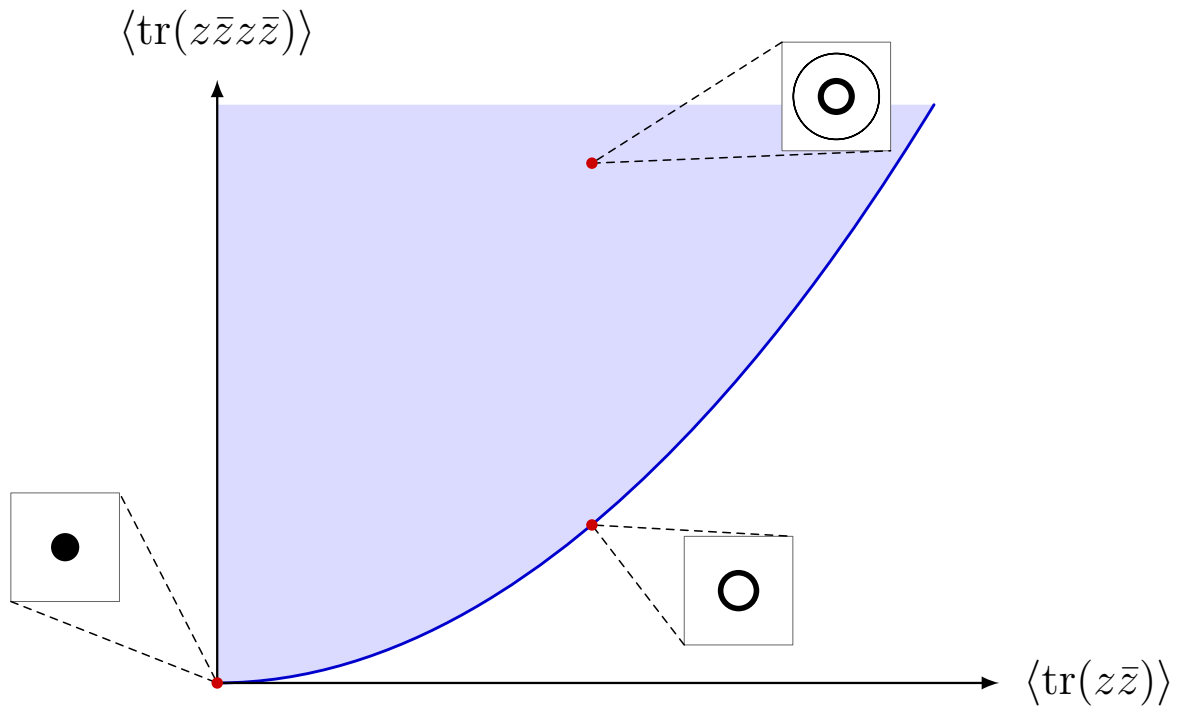

Of course, the $1/2$-BPS sector can be understood analytically in a multitude of ways, and there is certainly no need for bootstrapping. In the gravity picture, the heavy $1/2$-BPS states are described by the Lin-Lunin-Maldacena geometry \cite{Lin:2004nb}.  These gravity solutions are parametrized by a droplet density $\rho(\rz)$  that takes the value either $0$ or $1$ with $\rz \in \mathbb{C}$. Using $\rho(\rz)$, one can compute $\langle \tr{z\bar{z}} \rangle$ and $\langle \tr{z\bar{z}z \bar{z}} \rangle$ by\footnote{The derivation can be found in various places, such as \cite{Takayama:2005yq,Anempodistov:2025maj,Chen:2024oqv}. In comparison with Appendix A of \cite{Chen:2024oqv}, we have
$\langle \mathrm{tr}(z\bar z)\rangle_{\textrm{here}} = \langle \mathrm{Tr}(Z\bar Z)-N^2\rangle_{\textrm{there}}/N^2$
and
$\langle \mathrm{tr}(z\bar z z\bar z)\rangle_{\textrm{here}} = \langle \mathrm{Tr}(Z\bar Z Z\bar Z)-3N\mathrm{Tr}(Z\bar Z)+N^3\rangle_{\textrm{there}}/N^3$.
These differences arise because here $z,\bar z$ label creation and annihilation operators, while there $Z,\bar Z$ represent complex matrices in a Gaussian matrix integral.}
\begin{equation}
\begin{aligned}
	\langle \tr{z\bar{z}} \rangle & = \int \frac{\d^2 \rz}{\pi} \rho(\rz) \left(|\rz|^2 - \frac{1}{2}\right)\,,\\
	\langle \tr{z\bar{z}z\bar{z}} \rangle  & = \int \frac{\d^2 \rz}{\pi} \rho(\rz ) \left(|\rz|^4 -  |\rz|^2 + \frac{1}{6}\right)\,. 
\end{aligned}
\end{equation} 
If we focus on the first terms in the brackets, these equations have the simple interpretation of replacing $z$ by a diagonal matrix whose eigenvalue distribution is given by $\rho/\pi$. The other terms in the bracket are corrections that are subleading when the eigenvalues are large. It is simple to recover the inequality (\ref{z4z2}) using this perspective. 
Denote $u=\langle\tr{z\bar z}\rangle$, using
\begin{equation}
\int\frac{\d^2\rz}{\pi}\rho(\rz)=1\,,
\qquad
\int\frac{\d^2\rz}{\pi}\rho(\rz)|\rz|^2=u+\frac{1}{2}\,,
\end{equation}
we obtain
\begin{align}
\langle\tr{z\bar z z\bar z}\rangle-u^2
={}&\int\frac{\d^2\rz}{\pi}\rho(\rz)
  (|\rz|^2-u)(|\rz|^2-u-1)+\frac{1}{6}
  \notag\\
\geq{}&\int_0^1 \d t\,t(t-1)+\frac{1}{6}=0\,.
\end{align}
From the second to third line, we replaced $0\leq\rho \leq 1$ by one inside the annulus
$u\leq|\rz|^2\leq u+1$, where the polynomial is negative,
and zero outside of it. The bound is saturated when the droplet takes the shape of a single annulus, as shown in Figure \ref{fig:droplet}. 

The fact that (\ref{z4z2}) does not contain an upper bound for $\langle \tr{z\bar{z}z\bar{z}} \rangle$ has a simple gravity interpretation. Starting from the annulus configuration, one can increase $\langle \tr{z\bar{z}z\bar{z}} \rangle$ unboundedly by simply moving a tiny fraction of the droplet to a distant region, while keeping fixed $\langle \tr{z\bar{z}}\rangle$ by shrinking the original annulus slightly. In other words, the unbounded nature of the moduli space of gravity solutions maps directly to the unbounded direction in the bootstrap.

The $1/2$-BPS case serves as a simple demonstration of the bootstrap idea, but it is certainly more convenient to analyze using other techniques. On the other hand, for heavy states in the $1/4$- and $1/8$-BPS sectors, we do not have systematic methods to compute expectation values of operators directly in the large $N$ limit, so bootstrap already provides non-trivial new information. 

Let's consider the  $1/4$-BPS sector, built out of letters $x$ and $z$. Within this sector, the supercharge loop equation (\ref{Qloop}) can be equivalently cast as
\begin{equation}\label{quarterloop}
	\langle \tr{ [x,z] ... }\rangle = 0\, 
\end{equation}
where the dots denote any strings of $x,z$ and their conjugates. We will further restrict our attention to states that are singlet under $SU(2)_R \subset SO(6)_R$, under which $x,z$ transform as a doublet.

An interesting question in the $1/4$-BPS sector is whether the BPS states allow for an \emph{eigenvalue} description, similar to the $1/2$-BPS case. Indeed, as far as state counting is concerned, one can view $x$ and $z$ as simultaneously diagonalizable matrices \cite{Kinney:2005ej}.  
We expect such a commuting matrix picture to be valid in the regime where the occupation numbers of matrix oscillators are large. This happens in the limit where the $R$-symmetry charges are much greater than $N^2$, or, in terms of our normalized variables, $\langle \tr{z\bar{z}}\rangle, \langle \tr{x\bar{x}}\rangle \gg 1$. In this limit, as the bosonic oscillators are highly occupied, the dynamics of the system should become classical. The BPS states should describe the minimum of the classical potential $V \propto \Tr{[X,Z][\bar{Z},\bar{X}]}$, where the two matrices commute. Using bootstrap, we can verify this expectation and furthermore, constrain deviations away from it.

\begin{figure}[t!]
    \centering
    \includegraphics[width=\columnwidth]{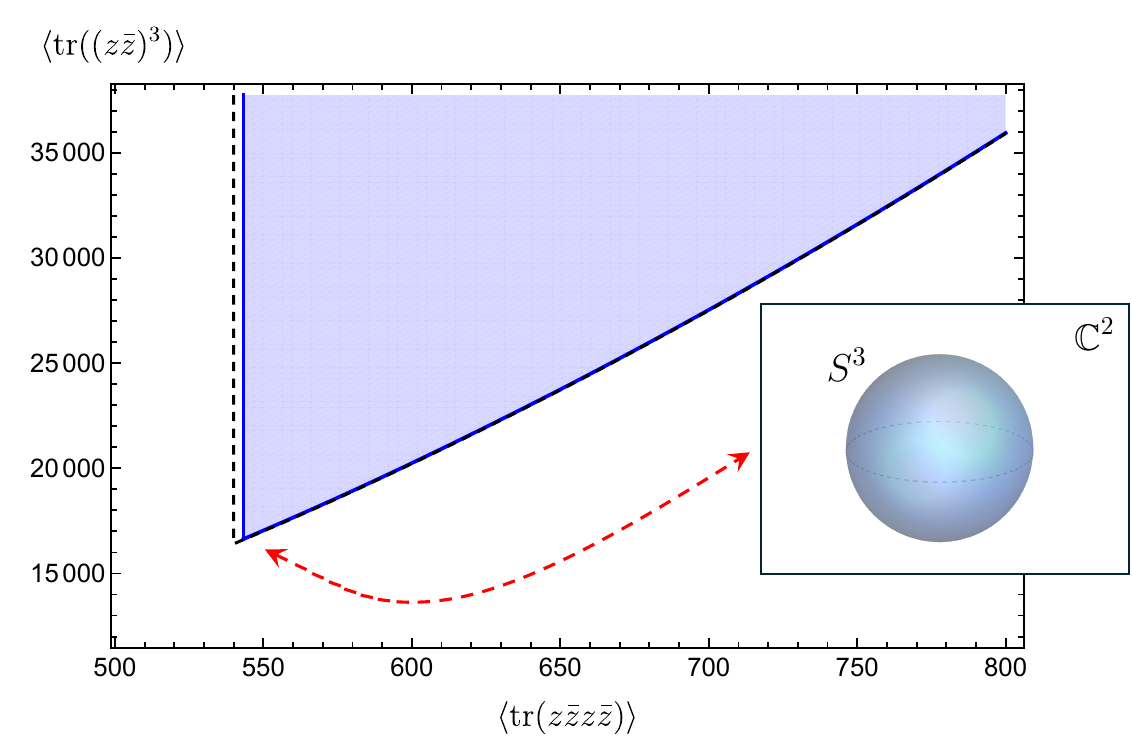}
    \caption{The allowable region in the $\langle\tr{(z\bar{z})^2}\rangle$ - $\langle\tr{(z\bar{z})^3}\rangle$ plane from the level $4$ bootstrap (blue), with the level $3$ boundary shown as a black dashed line (see Appendix \ref{app:bootstrap} for how levels are defined). The bottom corner of the blue region agrees well with an $S^3$ eigenvalue density.}
    \label{fig:quartercompare}
\end{figure}

In Fig. \ref{fig:quartercompare}, we  fix $\langle \tr{z\bar{z}}\rangle  = \langle \tr{x\bar{x}}\rangle = 20\gg 1$ and plot the feasible region in $\tr{ (z\bar{z})^3 }$ versus $\tr{(z\bar{z})^2}$. 
Similar to the half BPS case, one gets a continent that is unbounded in both directions. 
It is interesting to focus on the lower left corner, which should be compared to a configuration where the eigenvalues are most densely packed, if such a description is valid. A natural guess for an $SU(2)_R$ invariant eigenvalue density distribution would be
\begin{equation}\label{quarterdensity}
	\rho(\rx,\rz) =  \frac{1}{\pi^2 r_0^2} \,\delta\left(|\rx|^2 + |\rz|^2 - r_0^2 \right)
\end{equation}
which describes an $S^3$ with radius $r_0$ inside $\mathbb{C}^2$ \cite{Berenstein:2005aa}. Replacing $z,\bar{z},x,\bar{x}$ by diagonal matrices, one can estimate various moments using $\rho$.
In Table \ref{tab:shellcomparison}, we compare the expectation values of various observables at the level $4$ corner point, computed from bootstrap, with those moments computed from $\rho$ (we fix $r_0^2 = 40$ using the value of $\langle \tr{z\bar{z}}\rangle$).  From the table, we see that the joint eigenvalue density (\ref{quarterdensity}) provides a relatively accurate characterization of the observables. If we interpret the table in the opposite direction, it also suggests that our bootstrap bounds have converged to the actual bounds reasonably well.

\begin{table}[h!]
    \centering
    \setlength{\tabcolsep}{7.5pt}
    \begin{tabular}{cccc}
        \hline
        Observable & Bootstrap & $\rho(\rx,\rz)$ & Deviation \\
        \hline
        $\langle\tr{z\bar z z\bar z}\rangle$
            & 543.32 & 533.33 & $+1.87\%$ \\
        $\langle\tr{zx\bar z\bar x}\rangle$
            & 268.32 & 266.67 & $+0.62\%$ \\
        $\langle\tr{z\bar z x\bar x}\rangle$
            & 261.66 & 266.67 & $-1.88\%$ \\
        $\langle\tr{(z\bar z)^3}\rangle$
            & 16639.20 & 16000.00 & $+3.99\%$ \\
        $\langle\tr{z\bar z z\bar z x\bar x}\rangle$
            & 5253.11 & 5333.33 & $-1.50\%$ \\
        \hline
    \end{tabular}
    \caption{ We compare the moments at the bottom point of the level $4$ boundary in Fig. \ref{fig:quartercompare} with those computed from $\rho(\rx,\rz)$. Note that among the five rows presented, only three are independent, see Appendix \ref{app:quarter}.}
    \label{tab:shellcomparison}
\end{table}

When $\langle \tr{z\bar{z}}\rangle$ is of order one, a naive eigenvalue description no longer applies, but bootstrap still yields non-trivial bounds. It would be interesting to compare such bounds with computations using the coherent state representation of $1/4$-BPS states \cite{Berenstein:2022srd,Lin:2022wdr,Holguin:2023naq}. Using such a representation, expectation values of single trace operators can be reduced to a unitary matrix integral, though they do not appear easy to analyze in the large $N$ limit. One could also try to compare the bootstrap results with exact diagonalization results \cite{McLoughlin:2020zew,Chen:2024oqv} in the quarter BPS sector. It would also be interesting to interpret the bootstrap results 
in terms of the $1/4$-BPS bubbling geometries in gravity (see \cite{Liu:2004ru,Chen:2007du} and references therein). Of course, one should keep in mind that our  bootstrap is only set up at weak couplings, while observables we bound are generally coupling dependent.

We leave a similar analysis in the $1/8$-BPS sector to Appendix \ref{app:quarter}, in which case the eigenvalue distribution in (\ref{quarterdensity}) is replaced by an $S^5 \in \mathbb{C}^3$.

\section{The one-loop BMN Sector}

After warming up with simpler sectors that only contain BPS states that  are known as monotone or graviton states, let's turn to the full BMN sector that is known to contain fortuitous, or non-graviton states with order $N^2$ entropy \cite{Choi:2023vdm}. We will refrain from calling them black hole states for the moment, since after all, it will be the task of the bootstrap to establish that they are really black hole-like.

The one-loop BMN sector has an $SU(3)_{R} \subset SO(6)_R$ global symmetry, where the scalars $\phi^1,\phi^2,\phi^3$ and fermions  $\psi_1,\psi_2,\psi_3$ transform as fundamental and anti-fundamental, respectively. In the following, we focus on BPS states that are invariant under $SU(3)_{R}$ transformations, leading to additional loop equations
\begin{equation}
	\langle [J_{ij}, \mathcal{O}]\rangle = 0 \,,\quad \forall \,\mathcal{O},
\end{equation}
where $J_{ij}$ are the generators of $SU(3)_R$ (see Appendix \ref{app:BMN} for more detail). As a consequence, all three normalized $R$-charges $R_i/N^2$ become equal, which we denote by $r$.  Furthermore, the BMN truncation automatically satisfies $J_1 = J_2$, and we denote their large $N$ normalized values as $j$. Therefore, the states we consider can be characterized by two charges $(r,j)$, which can be further
 expressed in terms of the ``number operators" of the scalars and fermions 
\begin{equation}\label{rj}
\begin{aligned}
r &= \frac{1}{3}n_{\phi}+\frac{1}{6}n_{\psi}\,,\quad
j = \frac{1}{2}n_{\psi}+n_f\,.
\end{aligned}
\end{equation}
Here $n_{\phi} = 3\langle \tr{\phi^1 \bar{\phi}^1} \rangle$, $n_{\psi} =3 \langle \tr{\psi_1 \bar{\psi}_1} \rangle$ and $n_f = \langle \tr{f \bar{f} } \rangle$. We have $n_\phi,n_f\geq0$ while $0\leq n_\psi\leq 3$.

%

A very first question one typically asks when studying BPS states is how they are distributed. In our context, we can ask which part of the $(r,j)$ plane is allowed to host BPS states, including both the multi-graviton and non-graviton states. For the multi-graviton states, we already know from the explicit construction of their cohomologies \cite{Chang:2013fba} that they must reside in the region $j\leq 3r$, see Appendix \ref{app:BMN}.\footnote{Here we only consider the cohomologies in the $SU(N)$ theory. The $U(N)$ theory contains additional cohomologies such as powers of the trace mode $\tr{f}$ that would violate $j\leq 3r$. Even though our large $N$ loop equations are insensitive to the differences between $SU(N)$ and $U(N)$, we've imposed $\langle \tr{f}\tr{\bar{f}}\rangle =0$ so the trace mode cannot have order $N^2$ occupancy.}
On the other hand, we have less of an idea about where the non-graviton states are distributed, except from explicit finite $N$ examples \cite{Choi:2023znd,Choi:2023vdm,deMelloKoch:2024pcs,Gaikwad:2025ugk,Kim:2026rnx,deMelloKoch:2026sbd,Kibe:2026qid,Kim:2026mch,deMelloKoch:2026cvn,Benjamin:2026FortuityAI}.\footnote{Many of the recent results concern non-graviton states that are outside the $SU(3)_R$ singlet sector we focus on. We leave a detailed comparison to future work.} A motivating question, which we do not resolve here, is whether the non-graviton states satisfy a concentration property of \cite{Chen:2024oqv}, under which they only reside along a one dimensional line in the $(r,j)$ plane.\footnote{One can assume concentration and try to infer the locus from the index. To our knowledge, the index of the BMN sector is not well-understood away from the small charge regime \cite{Choi:2023vdm,Chang:2024lkw,Zigdon:2025icw}.}

As a proof of concept that bootstrap is in principle capable of addressing such questions, we can apply similar bootstrap ideas to the case of $N=2$ (see Appendix \ref{app:bootstrap} for more comments). The bootstrap rules out much of the $(R,J)$ space and the allowed region localizes around the known families of fortuitous operators. This suggests that the enumeration of operators in \cite{Choi:2023znd,Gaikwad:2025ugk,Kim:2026mch} already provides a complete picture at $N=2$, as was strongly suggested but not proven by the index.
\begin{figure}[t!]
    \centering
    \includegraphics[width=0.9\columnwidth]{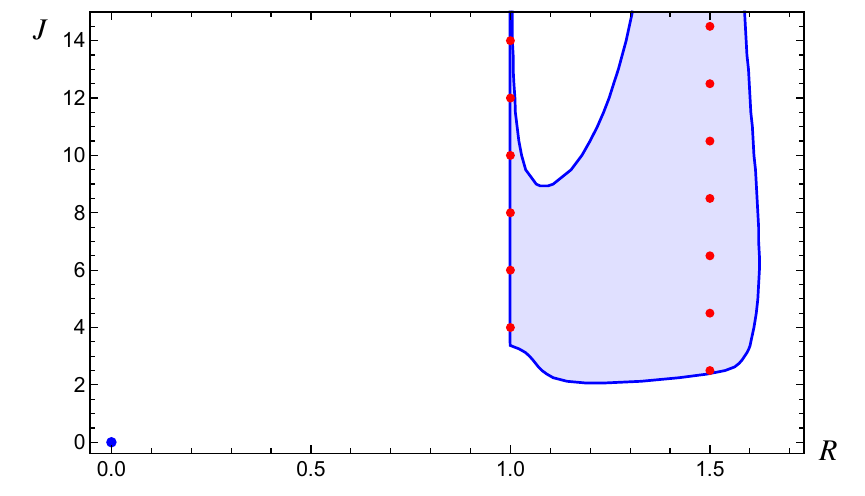}
    \caption{When the bootstrap method is applied at $N=2$, the allowable region ({\color{blue}{blue}}) tightly wraps known towers of non-graviton operators ({\color{red}{red}} dots). Note that apart from the vacuum state at $(0,0)$, there are no $SU(3)_R$ singlet graviton states at $N=2$ \cite{Gaikwad:2025ugk}.}
    \label{fig:bmn_n2}
\end{figure}

\begin{figure*}[t!]
    \centering
    \subfloat[\label{fig:bmn_gamma}]{\includegraphics[width=0.45\textwidth]{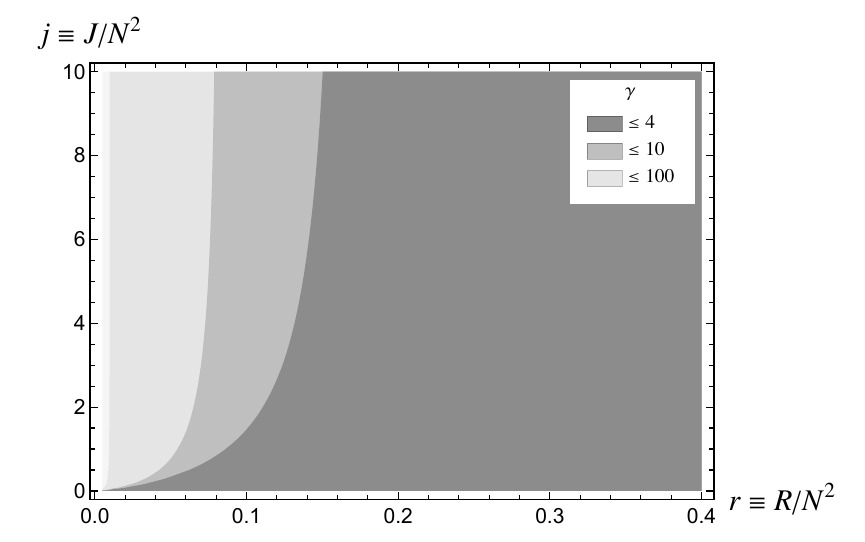}}\hfill
    \subfloat[\label{fig:bmn_composition}]{\includegraphics[width=0.45\textwidth]{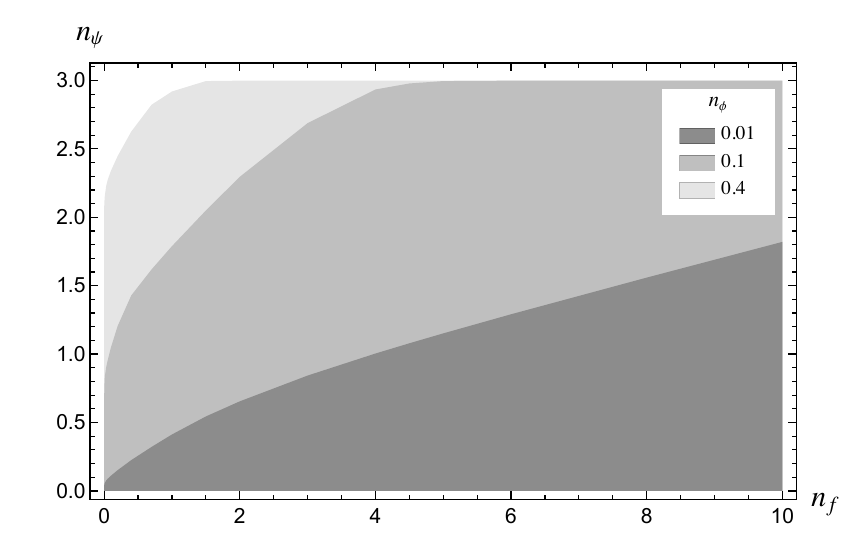}}
    \caption{(a) We show regions in the $r$-$j$ plane where the minimal ratio $\gamma$ between the quartic moment $\langle \tr{f\bar{f}f\bar{f}}\rangle$ and $n_f^2$ can be smaller than various values. Towards the $r=0$ axis, $\gamma$ is forced to be large and the $r=0$ axis is rigorously ruled out. (b) We show the allowed region in the $n_f$-$n_\psi$ plane with different values of $n_\phi$.}
    \label{fig:bmn_large_n}
\end{figure*}

Returning to the large $N$ bootstrap, however, we have not found decisive evidence that any finite portion of the $r,j\geq 0$ quadrant is ruled out.  
Nonetheless, we see a curious feature that when approaching the $r=0$ axis with fixed $j$, the quartic moment of the field strength letter $f$, $\langle \tr{f\bar{f}f\bar{f}}\rangle$ must become increasingly large. We define a quantity
\begin{equation}
	\gamma \equiv \textrm{min}\frac{\langle \tr{f\bar{f}f\bar{f}}\rangle}{n_f^2}
\end{equation}
and in Fig. \ref{fig:bmn_gamma} we show the regions that are allowed under the assumption that $\gamma$ is smaller than different cutoffs. Note that a ``natural" scale for $\gamma$ would be order one, yet, approaching $r=0$ requires $\gamma$ to take diverging values. We suspect that this hints that at higher levels in the bootstrap, regions near $r=0$ with finite $j$ will be ruled out.\footnote{For comparison, among the known $SU(3)_R$ singlet non-graviton operators, the smallest value of $r = R/N^2$ is $2/9 \approx 0.22$, achieved by an operator at $(R,J)=(2,5)$ in the $N=3$ theory.}

To obtain some more information about the BPS states, we could explore how they are distributed in the three dimensional space of $n_\phi,n_\psi, n_f$. Even though these quantities are not individually conserved, they provide a simple picture of the ``composition" of the BPS states. In Fig. \ref{fig:bmn_composition}, we show the allowed region in the $n_f$ - $n_\psi$ plane for different choices of $n_\phi$. In particular, we see that as $n_\phi\rightarrow 0$, the entire $n_\psi$ axis becomes ruled out, suggesting that there are no BPS states made out of only the fermions $\psi_i$ (in the large $N$ sense).\footnote{I thank Siyul Lee for suggesting making such statements.}

Given that the large $N$ bootstrap sees both the graviton and non-graviton states, a natural question is whether there exists a set of operators that can tell them apart. Put differently, if indeed the non-graviton states correspond to black holes in the bulk, albeit in a very stringy regime, what operators can probe the existence of the horizon? 

\begin{figure}[b!]
    \centering
    \includegraphics[width=\columnwidth]{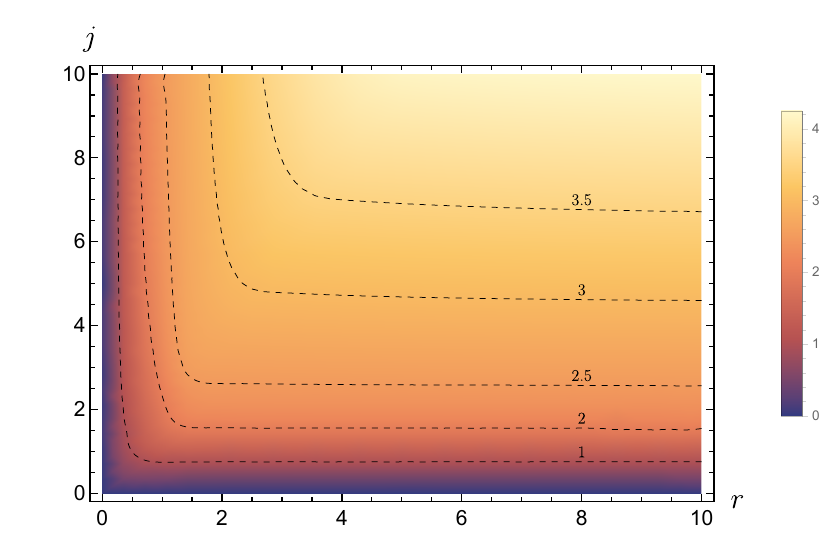}
    \caption{Heat map for the maximal value of $\mathcal{D}_\phi$ on the $(r,j)$ plane. At fixed $j$ and increasing $r$, $\textrm{max}\,\mathcal{D}_\phi$ first rises and then plateaus. This means that at $r\gg 1$, adding scalars no longer increases $\textrm{max}\,\mathcal{D}_\phi$.}
    \label{fig:bmn_scalar_comm}
\end{figure}

We have not yet found a satisfying answer to this question. We anticipate that going to higher levels in the bootstrap could lead to much tighter bounds that can help us identify such ``horizon indicators".\footnote{One might try to use the knowledge that the graviton states are restricted to $j\leq 3r$ to look for such indicators. Unfortunately, we have not identified a sharp difference across the $j=3r$ line. } Here instead, we will boldly proceed with our existing bounds and study a particular operator 
	$\mathcal{D_\phi} \equiv \frac{1}{3}\sum_{i<j} \langle\tr{[\phi^i, \phi^j][\bar{\phi}^j, \bar{\phi}^i]}\rangle$,
which measures the non-commutativity of the scalar matrices.\footnote{We denote it by $\mathcal{D}_\phi$ since it is proportional to the one-loop dilatation operator in the bosonic $1/8$-BPS sector. For this reason, $\mathcal{D}_\phi$ is forced to be zero when $j=0$, where only $n_\phi$ can be nonzero.} Our expectation is that black hole operators should have a considerable amount of $\mathcal{D}_\phi$, as compared to e.g. $n_\phi^2$. The motivation of this is twofold. First, a small value of $\mathcal{D}_\phi$ indicates that the scalar matrices $\phi^i$ approximately commute, which suggests a description of the scalar sector in terms of their eigenvalue distribution, which involves only an $\mathcal{O}(N)$ number  of degrees of freedom. On the other hand, one expects black hole states to involve all $\mathcal{O}(N^2)$ degrees of freedom. The second piece of motivation comes from the expectation that the non-graviton operators are ``non-Coulomb" \cite{Chang:2025mqp,Choi:2025bhi}, meaning that their cohomology classes become trivial if the matrices are taken to be diagonal, so they must make use of the non-diagonal parts of the matrices in a non-trivial way.\footnote{This argument is only heuristic since it is not entirely clear how Coulombness  is translated into quantum expectation values. Furthermore, it is recently found that the graviton states can also be non-Coulomb \cite{Kim:2026mch}.}

Interestingly, using simple bootstrap constraints, one can derive analytically an upper bound for $\mathcal{D}_\phi$ \cite{GPT6Astra:2026ScalarBound}
\begin{equation}\label{Dphi}
	\mathcal{D}_\phi \leq  4r .
\end{equation}
From (\ref{rj}), we know that $r$ depends linearly on $n_\phi$. Therefore, this means that in the regime $r\gg 1$, the commutator can \emph{never} be as big as a generic quartic moment, which would scale as $n_\phi^2$ instead. This implies that if our prejudice of black hole states having large $\mathcal{D}_\phi$ is correct, then they cannot be present at $r\gg 1$. We can nonetheless have states that are black holes dressed by graviton operators that carry $r\gg 1$ that does not contribute to $\mathcal{D}_\phi$. An example of such operators is the dual giant dressed black hole operators discussed in \cite{deMelloKoch:2024pcs}.

In Fig. \ref{fig:bmn_scalar_comm}, we plot the maximal value of $\mathcal{D}_\phi$ that is allowed by the numerical bootstrap on the $(r,j)$ plane.\footnote{Note that these bounds are rigorous, but they are rather loose, since to achieve them, some higher moments (length six variables) need to tend to infinity.} The numerical bound strengthens the bound (\ref{Dphi}) by showing that the maximal value of $\mathcal{D}_\phi$ plateaus as one increases $r$. Guided by this, we conjecture that the operators corresponding to black holes (not dressed by further graviton operators) are concentrated in regions where $r$ is of order one.

\section{Discussion}

In this paper, we explored the matrix bootstrap method as a way to access properties of heavy BPS states with scaling dimensions $\mathcal{O}(N^2)$ directly in the large $N$ limit. We explain the distinction that is expected for horizonless geometries that carry explicit moduli in the large $N$ limit versus supersymmetric black holes in the bootstrap program. Assuming infinite power of the bootstrap, after all the conserved charges are fixed, one expects to find a continent for the horizonless geometries from their moduli, while for the black hole one expects a converging island, as predicted by the Schwarzian theory. 

We implemented the bootstrap program for the one-loop BMN sector of $\mathcal{N}=4$ SYM. For special states that are $1/2, 1/4$ and $1/8$-BPS, bootstrap agrees with their eigenvalue pictures in the large charge limit while rigorously bounding deviations at smaller charges. For the $1/16$-BPS states, we discussed what one can already infer from our numerical results, though challenges remain to isolate the black hole states cleanly from the multi-graviton states.

There are many technical aspects of our work that can be improved. It is particularly interesting to make further connections with explicit finite $N$ constructions. We close with a few other questions that we think would be interesting to investigate.

\setlength{\leftmargini}{1em}

\begin{itemize}
\item  	\emph{Strong couplings.} Our exploration has been limited to weak couplings in the field theory, where the bulk is highly stringy. In matrix quantum mechanics systems, it is straightforward to go to strong couplings. One expects to find that operators corresponding to massive string modes should have expectation values that are suppressed compared to those of supergravity. It would be nice to see this explicitly. 
\item \emph{Finite $N$ bootstrap and fortuity}. In the finite $N$ bootstrap, one can simply treat $N$ as a formal parameter that enters the loop equations. Fortuity should manifest as new allowable regions that pop up when $N$ becomes small. It would be interesting to see this.
\item \emph{The Schwarzian gap.} A natural target for the finite $N$ bootstrap would be to put a bound on the gap above the BPS states. Schwarzian theory \cite{Stanford:2017thb,Boruch:2022tno} predicts the size of the gap is of order $1/N^2$. It would be nice to establish this rigorously (see also \cite{Chang:2023zqk,Cabo-Bizet:2024gny}).

\item \emph{Chaos versus constraints.} It is interesting to ask how BPS chaos emerges in the BPS bootstrap. Our expectation of the continent versus island suggests that chaos can be linked to how stringent the bootstrap constraints are.  It would be nice to explore this further in simpler examples.
\end{itemize}

\section*{Acknowledgments}

We acknowledge helpful conversations with Song Bao, Chi-Ming Chang, Simon Caron-Huot, Minjae Cho, Tom Hartman, Ji Hoon Lee, Siyul Lee, Henry Lin, Brian McPeak, Harish Murali and Zechuan Zheng. We are grateful to Minjae Cho and Henry Lin for comments on a draft.  We acknowledge the ``LITP Workshop on Quantum Black Holes" at the University of Michigan in March 2026, where some preliminary results were presented. 

\textbf{AI Disclosure:} The coding involved in this paper is performed by Codex (except some basic demo codes written by the author). We acknowledge the OpenAI researcher program and Mary Xia for providing free access to ChatGPT Pro accounts.

%
%
%
 \appendix

 \setcounter{secnumdepth}{3}
 \section{Additional details of the bootstrap}
 \label{app:bootstrap}

 In this appendix, we mention some additional details of the bootstrap program that were omitted in the main text. For the most part, as one could examine from the published solver \cite{BPSBootstrapCode}, we follow the approach in \cite{Lin:2024vvg,Lin:2025srf} closely, to which we refer our readers for detailed explanations of the bootstrap approach. However, a difference is that even though we've imposed various symmetry constraints as loop equations, we did not take them into account when building the moment matrix, which \cite{Lin:2024vvg,Lin:2025srf} had done systematically. This leaves room for further improvement. 
 
 In the bootstrap program, we've made use of the fact that if $\mathcal{M}\succeq 0$ then $\textrm{Re}(\mathcal{M}) \succeq 0$ to set Hermitian conjugate observables real and equal.  Apart from imposing $\mathcal{M}\succeq 0$, we have also implemented a multi-trace relaxation procedure \cite{Lin:2025srf} up to double trace operators. In the bootstrap scans, if there are quantities that are held fixed (such as $\langle\textrm{tr}(z\bar{z})\rangle$ in Fig. \ref{fig:quartercompare}), we use large $N$ factorization to substitute such factors in multi-trace operators by $c$-numbers.
 
 We define the level of the bootstrap $\ell$ to be half of the longest length of single traces that are included in the moment matrix $\mathcal{M}$ (all letters count as length one). For the $1/4$-BPS sector results shown in the paper, apart from the level $3$ comparison in Fig. \ref{fig:quartercompare}, we've used $\ell=4$, meaning that we kept moments up to length eight. For the $1/8$-BPS and the BMN sector results (including the $N=2$ case), we used $\ell=3$ instead.

In Table \ref{tab:bootstrap}, we document some of the relevant scales of the bootstrap problem involved in this project. All of the computations in this paper were performed on a 2021 MacBook Pro, so we anticipate large room for improvement.
\begin{table}[h!]
\centering
\small
\setlength{\tabcolsep}{4pt}
\renewcommand{\arraystretch}{1.15}
\begin{tabular}{|c|ccc|ccc|}
\hline
\multirow{2}{*}{$\ell$}
& \multicolumn{3}{c|}{$1/4$-BPS}
& \multicolumn{3}{c|}{BMN} \\
\cline{2-7}
& $n_{\mathrm{free}}$ & $d_{\max}$ & $d_{\mathrm{total}}$
& $n_{\mathrm{free}}$ & $d_{\max}$ & $d_{\mathrm{total}}$ \\
\hline
2 &  2 &  4 &  13 &  27 &  6 &  58 \\
3 &  5 &  8 &  40 & 561 & 29 & 559 \\
4 & 14 & 24 & 151 & --  & -- & --  \\
\hline
\end{tabular}
\caption{The number of free variables $(n_{\textrm{free}})$, the maximal dimension ($d_{\rm max}$) and the total dimension ($d_{\rm total}$) of the positivity blocks in the bootstrap problems at different $\ell$.}
\label{tab:bootstrap}
\end{table}

In Fig. \ref{fig:bmn_n2} specifically, we've implemented the bootstrap at finite $N$ ($N=2$). A few things are different here. First, there is no longer large $N$ factorization, which increases the number of variables. However, we utilize the trace relations at $N=2$, which greatly reduces the number of variables. We also project strictly into $SU(3)_R$ singlet states by setting the Casimir to be zero. This is stronger than the large $N$ implementation, which can in principle allow a uniform mixed state of non-trivial representations of $SU(3)_R$. We further demand $\langle \mathcal{Q} \mathcal{O}\rangle = \langle \mathcal{O} \mathcal{Q}\rangle = 0$.

 \setcounter{secnumdepth}{2}
 \section{Additional details of the BMN sector}
 \label{app:BMN}

In this appendix, we provide some additional details on the BMN sector that are suppressed in the main text.

\paragraph{Charges and Symmetries} 
In the table below, we list the letters and their respective charges in the BMN sector.
\begin{table}[h!]
\centering
\renewcommand{\arraystretch}{1.4}
\begin{tabular}{c|c|c|c|c|c||c}
 & $R_1$ & $R_2$ & $R_3$ & $J_1$ & $J_2$ & $\Delta$ \\ \hline
$\phi^1$ & $1$ & $0$ & $0$ & $0$ & $0$ & $1$ \\ \hline
$\phi^2$ & $0$ & $1$ & $0$ & $0$ & $0$ & $1$ \\ \hline
$\phi^3$ & $0$ & $0$ & $1$ & $0$ & $0$ & $1$ \\ \hline
$\psi_1$ & $-\frac{1}{2}$ & $\frac{1}{2}$ & $\frac{1}{2}$ & $\frac{1}{2}$ & $\frac{1}{2}$ & $\frac{3}{2}$ \\ \hline
$\psi_2$ & $\frac{1}{2}$ & $-\frac{1}{2}$ & $\frac{1}{2}$ & $\frac{1}{2}$ & $\frac{1}{2}$ & $\frac{3}{2}$ \\ \hline
$\psi_3$ & $\frac{1}{2}$ & $\frac{1}{2}$ & $-\frac{1}{2}$ & $\frac{1}{2}$ & $\frac{1}{2}$ & $\frac{3}{2}$ \\ \hline
$f$ & $0$ & $0$ & $0$ & $1$ & $1$ & $2$ \\ \hline\hline
$\mathcal{Q}$ & $\frac{1}{2}$ & $\frac{1}{2}$ & $\frac{1}{2}$ & $-\frac{1}{2}$ & $-\frac{1}{2}$ & $\frac{1}{2}$
\end{tabular}
\caption{The letters and their charges in the BMN sector. $\Delta$ denotes the conformal dimension of the letters. }
\label{tab:bmncharges}
\end{table}

\noindent
The BMN sector has an $SU(1|3)$ symmetry. The relevant symmetry that we made use of in the paper is an $SU(3)_R$ subgroup with positive root generators given by
\begin{equation}
    J_{ij}\propto
    \textrm{Tr}\left(
        \phi^i\bar{\phi}^{j}
        -\psi_j\bar{\psi}_i\right), \qquad 1\leq i<j\leq 3.
\end{equation}

\paragraph{Graviton operators}  The BMN sector contains a set of single trace graviton operators, whose products form the multi-graviton/monotone states. Following the notation of \cite{Choi:2023vdm}, they include
\begin{equation}
\begin{aligned}
(u_n)^{i_1\cdots i_n}
    &= {\rm tr}(\phi^{(i_1}\cdots\phi^{i_n)}) \\
{(v_n)^{i_1\cdots i_{n\!-\!1}}}_j
    &= {\rm tr}(\phi^{(i_1}\cdots\phi^{i_{n\!-\!1})}\psi_j)
       -\textrm{``trace"} \\
(w_n)^{i_1\cdots i_{n\!-\!1}}
    &= {\rm tr}\bigl(\phi^{(i_1\cdots i_{n\!-\!1})}f \\
    &+{\textstyle\frac{1}{2}}\epsilon^{jk(i_p}
       {\textstyle\sum_{p=1}^{n\!-\!1}}  \phi^{i_1}\cdots\phi^{i_{p\!-\!1}}\psi_j
       \phi^{i_{p\!+\!1}}\cdots\phi^{i_{n\!-\!1})}\psi_k\bigr)\,.
\end{aligned}
\end{equation}
where $n\geq 2$ and $``\textrm{trace}"$ means contracting $SU(3)_R$ indices.
What are shown are the cohomology representatives, not the exact wavefunctions. What's important for us is how the three types of operators contribute to $(R,J)$. Here $R = (R_1+ R_2 + R_3)/3$ and $J = J_1 = J_2$. As shown in Table \ref{tab:gravitoncharges}, all such operators have $\Delta J \leq 3 \Delta R$. Since all the multi-graviton operators are given by acting them on the vacuum state, we conclude that the multi-graviton operators can only reside in the region $j\leq 3r$ in the large $N$ limit.
\begin{table}[h!]
    \centering
    \setlength{\tabcolsep}{7.5pt}
    \renewcommand{\arraystretch}{1.4}
    \begin{tabular}{ccc}
        \hline
        Operator & $\Delta R$ & $\Delta J$ \\
        \hline
        $u_n$ & $\frac{n}{3}$ & $0$ \\
        $v_n$ & $\frac{2n-1}{6}$ & $\frac{1}{2}$ \\
        $w_n$ & $\frac{n-1}{3}$ & $1$ \\
        \hline
    \end{tabular}
    \caption{Contributions of the single-trace graviton operators to $(R,J)$.}
    \label{tab:gravitoncharges}
\end{table}

  \setcounter{secnumdepth}{3}
 \section{Additional results in the $\frac{1}{4}$- and $\frac{1}{8}$-BPS sectors}
\label{app:quarter}
 
\paragraph{Exact results from loop equations} Using the loop equations, we can derive exact relations among various moments. Here we exhibit some simple examples. 

Let's first study the effect of the $SU(2)_R$ transformation generated by $J_+ = \tr{z\bar{x}}$, under which
\begin{equation}
	\delta x=z,\qquad \delta\bar z=-\bar x,\qquad
\delta z=\delta\bar x=0.
\end{equation}
Since we focus on  $SU(2)_R$ invariant states, we have
\begin{equation}
	\begin{aligned}
0&=\langle\delta\tr{xz\bar z\bar z}\rangle \\
& =\langle\tr{zz\bar z\bar z}\rangle
   -\langle\tr{xz\bar x\bar z}\rangle-\langle\tr{xz\bar z\bar x}\rangle.
\end{aligned}
\end{equation}
Similarly, varying $\tr{x\bar z z\bar z}$ gives
\begin{equation}
	\begin{aligned}
0&=\langle\delta\tr{x\bar z z\bar z}\rangle\\
 &=\langle\tr{z\bar z z\bar z}\rangle
   -\langle\tr{x\bar x z\bar z}\rangle-\langle\tr{x\bar z z\bar x}\rangle.
\end{aligned}
\end{equation}
The BPS loop equation (\ref{quarterloop}) and its conjugate give further relations
\begin{equation}
	\begin{aligned}
\langle\tr{[x,z]\bar z\bar x}\rangle=0, \quad 
\langle\tr{xz[\bar x,\bar z]}\rangle=0.
\end{aligned}
\end{equation}
Finally, we also use the $SU(N)$ singlet constraint where
\begin{equation}
	\langle \tr{\mathcal{G} \mathcal{O}}\rangle = 0, \quad \mathcal{G} =[x,\bar x]+[z,\bar z]+2\,\mathbf{1} 
\end{equation}
Inserting $\bar z z$ and $z\bar z$ as $\mathcal{O}$, we get
\begin{equation}
\begin{aligned}
&3\langle\tr{zx\bar z\bar x}\rangle
   +\langle\tr{z\bar z}\rangle-2\langle\tr{z\bar z z\bar z}\rangle
   +\langle\tr{z\bar z x\bar x}\rangle = 0,\\
 &  \langle\tr{z\bar z z\bar z}\rangle
   +\langle\tr{z\bar z x\bar x}\rangle
   -3\langle\tr{zx\bar z\bar x}\rangle  = 0.
\end{aligned}
\end{equation}
Using these equations, one can reduce all the length four variables which are nonzero down to two free variables, $\langle\tr{z\bar z}\rangle$ and $\langle\tr{z\bar z z\bar z}\rangle$. For example the second and third lines in Table \ref{tab:shellcomparison} are simply related to the first line by
\begin{equation}\label{eq:quarterquarticrelations}
\begin{aligned}
\langle\tr{zx\bar z\bar x}\rangle
    &=\frac12\langle\tr{z\bar z z\bar z}\rangle
      -\frac16\langle\tr{z\bar z}\rangle\,,\\
\langle\tr{z\bar z x\bar x}\rangle
    &=\frac12\langle\tr{z\bar z z\bar z}\rangle
      -\frac12\langle\tr{z\bar z}\rangle\,.
\end{aligned}
\end{equation}
We see that $\langle\tr{zx\bar z\bar x}\rangle \neq \langle\tr{z\bar z x\bar x}\rangle$. If we naively try to compute them using an eigenvalue distribution $\rho(\rx,\rz)$, they would be translated to the same expression $\int \d^2 \rz \d^2 \rx \,\rho(\rx,\rz)|\rx|^2 |\rz|^2$. Such a description can however become accurate in the large charge limit, since the difference  is proportional to $\langle\tr{z\bar z}\rangle$, which becomes subleading compared to $\langle\tr{z\bar z z\bar z}\rangle$.

\paragraph{Checking the eigenvalue picture for $1/8$-BPS states}

In the main text, we presented some simple checks of the bootstrap result against an eigenvalue distribution (\ref{quarterdensity}) that is $S^3 \in \mathbb{C}^2$. For $SU(3)_R$ invariant $1/8$-BPS states, the direct analogue would be an eigenvalue distribution that is an $S^5 \in \mathbb{C}^3$
\begin{equation}
	\rho(\rx, \ry,\rz) = \frac{2}{\pi^3 r_0^4} \delta (|\rx|^2+|\ry|^2+|\rz|^2-r_0^2)\,.
\end{equation}
To test it, we follow the same procedure as in the $1/4$-BPS case. We fix  $\langle\tr{x\bar x}\rangle=\langle\tr{y\bar y}\rangle=\langle\tr{z\bar z}\rangle=20$ and look at the allowable region of $\langle \tr{(z\bar z)^2}\rangle$ versus $\langle \tr{(z\bar z)^3}\rangle$, then focusing on the point where $\langle \tr{(z\bar z)^3}\rangle$ reaches a minimum. We then compare various moments at this point with the results computed from $\rho(\rx, \ry,\rz)$ (with $r_0^2$ fixed by the quadratic moments to be $60$). As one can see from Table \ref{tab:eighthshellcomparison}, they agree reasonably well. 

\begin{table}[h!]
    \centering
    \setlength{\tabcolsep}{7.5pt}
    \begin{tabular}{cccc}
        \hline
        Observable & Bootstrap & $\rho(\rx, \ry,\rz)$ & Deviation \\
        \hline
        $\langle\tr{z\bar z z\bar z}\rangle$
            & 610.00 & 600.00 & $+1.67\%$ \\
        $\langle\tr{z\bar z x\bar x}\rangle$
            & 295.00 & 300.00 & $-1.67\%$ \\
        $\langle\tr{(z\bar z)^3}\rangle$
            & 22188.37 & 21600.00 & $+2.72\%$ \\
        $\langle\tr{z\bar z z\bar z x\bar x}\rangle$
            & 7220.39 & 7200.00 & $+0.28\%$ \\
        $\langle\tr{z\bar z x\bar x y\bar y}\rangle$
            & 3234.37 & 3600.00 & $-10.16\%$ \\
        $\langle\tr{zxy\bar z\bar x\bar y}\rangle$
            & 3596.36 & 3600.00 & $-0.10\%$ \\
        $\langle\tr{zx\bar z y\bar x\bar y}\rangle$
            & 3556.03 & 3600.00 & $-1.22\%$ \\
        \hline
    \end{tabular}
    \caption{Comparison between the bootstrap results and the results from the eigenvalue density, at the minimum of $\langle\tr{(z\bar z)^3}\rangle$ with $\langle\tr{x\bar x}\rangle=\langle\tr{y\bar y}\rangle=\langle\tr{z\bar z}\rangle=20$. Among the seven rows shown, only four are independent (with $\langle z\bar{z}x\bar{x}\rangle, \langle\tr{zxy\bar z\bar x\bar y}\rangle, \langle\tr{zx\bar z y\bar x\bar y}\rangle$ determined by other variables).}
    \label{tab:eighthshellcomparison}
\end{table}

%
%
%

\bibliographystyle{apsrev4-1}
\bibliography{references}

\end{document}